# Spin Transfer Torque and Tunneling Magnetoresistance Dependences on the Finite Bias Voltages and Insulator Barrier Energy

Chun-Yeol You[a*], Jae-Ho Han[b] and Hyun-Woo Lee[b]

[a] Department of Physics, Inha University, Incheon 402-751, Korea
[b] PCTP and Department of Physics, POSTECH, Pohang 790-784, Korea

We investigate the dependence of perpendicular and parallel spin transfer torque (STT) and tunneling magnetoresistance (TMR) on the insulator barrier energy in the magnetic tunnel junction (MTJ). We employed single orbit tight binding model combined with the Keldysh non-equilibrium Green's function method in order to calculate the perpendicular and parallel STT, and TMR in MTJ with the finite bias voltages. The dependences of STT and TMR on the insulator barrier energy are calculated for the semi-infinite half metallic ferromagnetic electrodes. We find that perfect linear relation between the parallel STT and the tunneling current for the wide range of the insulator barrier energy. Furthermore, the TMR also depends on the insulator barrier energy, which contradicts to the Julliere's simple model.

# 1. Introduction

The spin transfer torque (STT) [1,2] and tunneling magnetoresistance (TMR) are the key technologies in the current magnetism research due to its potential application of STT-MRAM (magnetoresistive random access memory) [3]. The information writing mechanism of the STT-MRAM is so called current induced magnetization switching (CIMS) based on STT phenomena in magnetic tunneling junction (MTJ). Since the STT is angular momentum transfer by the spin polarized electron current, the system is far from equilibrium. Therefore, a rigorous non-equilibrium treatment such as Keldysh non-equilibrium Green's function methods must be employed [4,5,6]. Well established Keldysh non-equilibrium Green's function methods have been successfully adopted to explain recent experimental observations for STT in MgO based MTJ [7,8,9]. In this study, we calculate the dependence of STT and TMR on the insulator barrier energy height in the frame of non-equilibrium Keldysh Green's function method with the finite bias voltage [4,5,6,10]. Free electron, single orbit, tight binding model is used in our calculations for the simple cubic half metallic semi-infinite ferromagnetic electrodes. We find that the perpendicular (out-of-plane) STT is an even function of the bias voltage, and the lower barrier energy gives larger STT. Since the perpendicular STT is the same as the interlayer exchange coupling, roughly speaking, the magnitude exponentially decreases with the barrier height. The parallel (in-plane) STT is neither an even nor odd function of the bias voltage. The parallel STT also decreases with barrier height, but the dependence is not a simple exponential. However, we find a perfect linear relation between the parallel STT and the total tunneling current. TMR shows strong bias dependence even though we did not consider in-elastic scattering. The strong bias dependences are mainly due to the band shift. At the finite bias, the TMR depends on barrier height, which contradicts to the simple Julliere's model [11].

## 2. Keldysh non-equilibrium Green's function method

We briefly summarize the Keldysh non-equilibrium Green's function method for the STT calculations in our study. More details can be found elsewhere [5,10]. The schematic sketch of the trilayer structure is shown in Fig. 1 (a). The left and right electrodes are semi-infinite and finite $N$ (= 5) insulator barrier layer is placed between them. We assumed that the left (right) electrode is polarizer (switching) layer. The magnetization direction of the polarizer layer is placed in the $xz$-plane with angle $\theta$ from positive $z$-axis. The magnetization direction of the switching layer is parallel to the positive $z$-axis. From the semi-infinite electrodes, we calculate the surface Green's function and then each insulator layer is added by the Dyson equation [12,13,14]. We considered single-orbit tight-binding model with simple cubic structure, and two ferromagnetic electrodes are considered identical. The exchange energy, $\Delta_{EX}$, of the ferromagnetic layer is 0.7 eV, and the on-site energy of spin up (down) is 2.3 eV (3.0 eV). For the simplicity, the half metal ferromagnetic electrodes are examined, and the hopping energy, $t_{hop}$ = -0.5 eV, is fixed for all layers. The on-site energy of the insulator barrier layer, $U_{Ins}$, is varied from 3.5 to 4.5 eV. Corresponding barrier energy heights, $V_{Ins}$ (= $U_{Ins}$ − 6 $|t_{hop}|$), are 0.5~1.5 eV. The on-site energy in the inside of the insulator barrier decreases linearly as $U_{Ins}(i) = U_{Ins} - i V_{Bias}/N$ with the bias voltage $V_{Bias}$. With the finite bias voltage, the spin current at ($n$-1)-th layer becomes by [5,6]

$$\langle \mathbf{j}_{n-1} \rangle = \frac{1}{2} \sum_{k_\parallel} \int \frac{d\omega}{2\pi} \mathrm{Tr}\left\{ \left[ G_{RL}^+(\omega) T - G_{LR}^+(\omega) T^\dagger \right] \boldsymbol{\sigma} \right\}. \tag{1}$$

More details and the meanings of each symbol are described in Ref. [5]. The charge current is easily obtained by replacing $\boldsymbol{\sigma}/2$ with the unit matrix multiplied by $e/\hbar$. The spin current is directly related with STT by $\mathbf{T} = \langle \mathbf{j}_{Ins} \rangle$ at the interface between the insulator and switching layer for the semi-infinite switching layer. Therefore, we can obtain the parallel and perpendicular STT with the charge current. In our study, the pessimistic definition of TMR = $(R_{AP} - R_P)/R_{AP}$ is used, therefore the TMR of half metal is 100 % with small bias voltage which is corresponding to the infinite TMR in the optimistic definition.

## 3. Spin transfer torque and tunneling magnetoresistance for various insulator energies

The STT has two components, parallel (in-plane) and perpendicular (out-of-plane). In our coordinate system, parallel (perpendicular) STT is $T_x$ ($T_y$). It is well-known that the perpendicular STT is small in the whole metallic systems. In metallic systems, the whole Fermi surface contributes to the Brillouin zone integration, and it rapidly decays. However, it is quite different in the MTJ system, where the insulator layer exists between two electrodes. It has been theoretically predicted and experimentally confirmed that the perpendicular STT is comparable to the parallel one in the MTJ [6,7,8]. The difference between metallic system and MTJ is originated from the in-plane momentum, $k_\parallel$, integral in the Brillouin zone [15]. The insulator barrier acts as a $k_\parallel$ filter, so that $k_\parallel$'s around $\Gamma$ point mainly contribute to the integration in the in MTJ system. Furthermore, the perpendicular STT is important in practical spin dynamics. Since the perpendicular STT is an even function of the bias voltage in the symmetric MTJ [16], it prefers either parallel or anti-parallel states for both signs of bias voltages, and the effect increases for higher bias. Therefore, it is important to understand the details of spin dynamics. It may cause back-hopping, which is switching back after once current induced magnetization switching is occurred [17].

According to Slonczewski [16,18], the STT is related with the magneto-conductance coefficients, the *torkance* is given by

$$\frac{d\dot{\mathbf{S}}_R}{dV} = \frac{\hbar}{4e}(G_{++} - G_{--} + G_{+-} - G_{-+})\mathbf{s}_R \times (\mathbf{s}_R \times \mathbf{s}_L), \qquad (2)$$

so they strongly depends on the insulator barrier energy height. Therefore, it is easy to imagine that the STT also strongly depends on the insulator barrier energy. The insulator barrier energy of MgO is a fixed value for a bulk, but it is not true for thin film in real MTJ stack. The barrier energy can be tailored by deposition and annealing conditions [19,20]. The low resistance-area product (RA) is another important parameter for the real device applications, due to the impedance matching in STT-MRAM, and better data read rate, noise consideration in hard disk read head [21]. The RA mainly depends on the insulator barrier thickness and the energy height. They can be optimized by careful fabrication conditions. Therefore, the study of the STT and TMR dependence on the barrier energy height is an important research subject.

### 3.1 Tunneling magnetoresistance

Before we discuss about STT, let us discuss about tunneling current and TMR. We depicted the tunneling current as a function of the barrier energy height, $V_{Ins}$, for $V_{Bias}$ = 0.1 and 1.0 V in Fig. 2. The tunneling current are calculated for $\theta$ = 0, $\pi/2$, and $\pi$, respectively. In these calculations, we find the followings; While simple WKB approximation gives exponential dependence of the tunneling current on the $\sqrt{V_{Ins}}$, our results show slight deviation from the simple exponential dependence. We plotted log-log scale to compare with WKB approximation. Furthermore, the slope of the decay also depends on the bias voltage and $\theta$. For $V_{Bias} = 0.1$ V case, anti-parallel state ($\theta = \pi$) shows smaller tunneling current, that is, larger resistance than parallel state ($\theta = 0$). However, the results are reversed for the large bias ($V_{Bias} = 1.0$ V), it implies negative TMR (as shown in Fig. 4). The $V_{Ins}$ dependent TMR is shown in Fig. 3 for $V_{Bias} = 0.1$ and 1.0 V. The TMR is determined by spin polarized density of state in the framework of the Julliere's model [22], and it implies that the TMR is independent on the barrier energy height. However, our Keldysh non-equilibrium Green's function results show the barrier energy height dependence. For small $V_{Bias}$ (= 0.1 V), the dependence is weak. However, it is more serious for large $V_{Bias}$ (= 1.0 V) as shown in Fig. 3. Even the sign of the TMR is changed for larger $V_{Bias}$ in our results. It must be pointed out that the TMR is measured with the finite bias voltage in the real device operations. The bias dependences of TMR are also plotted in Fig. 4 for selected $V_{Ins}$ (= 0.5~1.5 eV). It is widely accepted that the decrease of the TMR with bias voltage is mainly due to the magnon excitation. However, the magnon excitation is not considered in our calculation. Therefore, the decrease of the TMR ascribes the band shifts with the bias voltage. At zero bias, we obtain 100 % TMR (pessimistic) due to the half-metallic nature of the ferromagnetic layers. As we already mentioned, present results show much more complex behavior than simple Julliere's model. The Julliere's model is failed to describe the huge TMR in MgO based MTJ. The huge TMR in MgO based MTJ can be explained by the in-plane momentum conservation due to the epitaxial structure with the band symmetry selected tunneling [23,24,25,26]. Even though we do not consider the band symmetry selected tunneling to mimic MgO based TMR, we can conclude that the Julliere's model is too simple to describe the correct TMR with finite bias voltage and the barrier energy height must be considered in TMR study.

## 3.2 Perpendicular and parallel STT

Next, the perpendicular and parallel STT are considered. Since the STT is closely related with the spin dependent conductance, it must be sensitive on the barrier energy height. The dependence of perpendicular ($T_y$) and parallel ($T_x$) on the $V_{Ins}$ are depicted in Fig. 5 (a) and (b) for $V_{Bias}$ = 0.1 and 1.0 V, respectively. All STT results are calculated for $\theta = \pi/2$. The overall behavior of perpendicular STT is exponential decay with $V_{Ins}$. Especially, $V_{Bias}$ = 0.1 V looks like perfect exponential decay. However, more careful analysis reveals that there is a small deviation. Somewhat large deviation is found in small $V_{Ins}$ region for $V_{Bias}$ = 1.0 V case. In this region, the perpendicular STT is negative and we skipped negative values in order to use log-scale plot. The exponential dependences can be easily explained by the relation between the STT and spin dependent conductance. Especially, since the perpendicular STT is the interlayer exchange coupling, the exponential dependence is natural. However, the origin of the slight deviation from the exponential dependence is not clear. We also performed the same calculations with non-half-metal ferromagnetic layers, and stronger deviations from the exponential dependence are found (now shown here). It must be pointed out that the magnitude of the STT is very sensitive on the barrier energy height, lower barrier gives larger STT for a given bias voltage. Therefore, the lower barrier energy height guarantees more effective current induced magnetization switching.

The bias dependences of the perpendicular and parallel STT for selected $V_{Ins}$ (= 0.5~1.5 eV) are shown in Fig. 6 (a) and (b). Here, we also skipped the negative STT for log-scale plots. The perpendicular STT is even function of the $V_{Bias}$ for all $V_{Ins}$ by the symmetry. However, the parallel STT is neither even nor odd function as Theodonis explained [6].

Figure 7 (a) and (b) shows the perpendicular and parallel STT as a function of the current for a given $V_{Bias}$ = 0.1 and 1.0 V. The current values for each $V_{Ins}$ ($\theta = \pi/2$) in Fig. 2 are used as abscissa data with the corresponding perpendicular and parallel STT. We find perfect linearity between the parallel STT ($T_x$) and the total charge current. Surprisingly, all data points of $T_x$ for $V_{Bias}$ = 0.1 and 1.0 V are falling into the single linear curve in spite of the wide range of the current values. However, the perpendicular STT ($T_y$) shows large deviation from the linear relations. The perfect linear relation of the $T_x$ is associated with the Eq. (2), while the perpendicular STT has more complicate relation. In parallel STT case, only the states between left and right electrode Fermi energies contribute, which is also so for the tunneling current. However, the perpendicular STT is contributed by all occupied states from the bottom of the energy band to the Fermi energy.

It must be emphasized that the parallel STT is directly related with the tunneling current, which are slightly deviated from the perfect exponential dependence on $V_{Ins}$ the Fig. 2. Therefore, the total current and parallel STT are not perfect exponential function of $V_{Ins}$, but they have linear relation between themselves. It may give a clue to the question: which one is more fundamental driving force of the STT between current and voltage in MTJ?

## 4. Conclusion

We investigate the TMR and STT in the frame of the non-equilibrium Green's function method for the symmetric MTJ. We varied the insulator barrier energy height. We found that TMR shows more complicate bias voltage and barrier height dependence behaviors then simple Julliere's model. Furthermore, the perpendicular and parallel STT shows approximately exponential decay with barrier energy height with somewhat un-expected deviations. However, we found a perfect linear relation between parallel STT and the total tunneling current.

## Acknowledgements

This work is supported by Nano R&D (2008-02553) and Mid-career Researcher Program (2010-0014109) programs through the NRF grant funded by MEST.

## Figure Captions

**Fig. 1.** (a) Schematic diagram of the MTJ layer structure. The semi-infinite left and right ferromagnetic leads are connected with insulator layer. The direction of the magnetization of polarizer layer is placed in the xz-plane with angle $\theta$ from positive z-axis. The magnetization direction of switching layer is parallel to the positive z-axis. (b) Band structures of the ferromagnetic and insulator layers.

**Fig. 2.** Tunneling current ($\theta = 0$, $\pi/2$, $\pi$) as a function of $\sqrt{V_{Ins}}$ for $V_{Bias} = 0.1$ and 1.0 V. In order to compare with WKB approximation, we used log-log plot.

**Fig. 3.** TMR (pessimistic) as a function of $V_{Ins}$ for $V_{Bias} = 0.1$ and 1.0 V.

**Fig. 4.** Bias dependent TMR (pessimistic) as a function of $V_{Bias} = 0.1$ and 1.0 V for various $V_{Ins}$ ($= 0.5 \sim 1.5$ eV).

**Fig. 5.** (a) Perpendicular ($T_y$) and (b) parallel ($T_x$) STT as a function of $V_{Ins}$ for $V_{Bias} = 0.1$ and 1.0 V.

**Fig. 6.** (a) Perpendicular ($T_y$) and (b) parallel ($T_x$) STT as a function of $V_{Ins}$ for $V_{Bias} = 0.1$ and 1.0 V. (The negative values are skipped for the log-scale plots).

**Fig. 7.** Perpendicular ($T_y$) and (b) parallel ($T_x$) STT as a function of total current. The current values from Fig. 2 ($\theta = \pi/2$) for various $V_{Ins}$ with fixed $V_{Bias}$, and the corresponding STT are plotted.

**Fig. 1.**

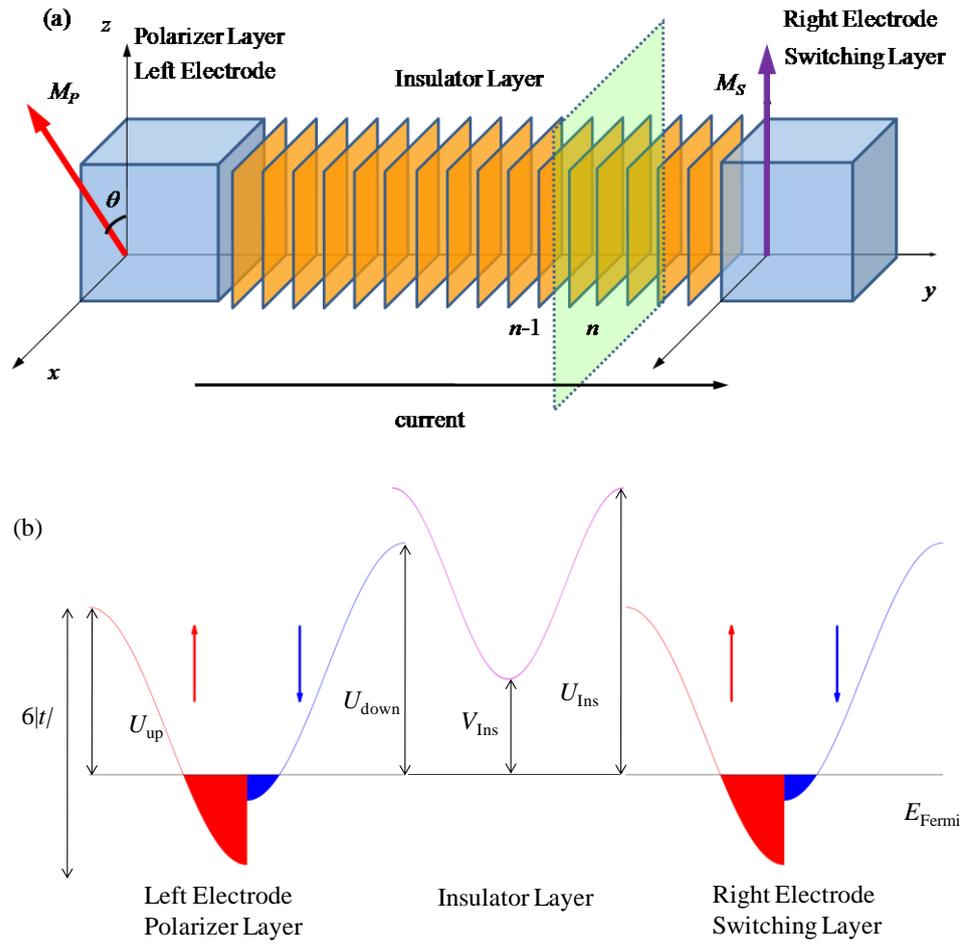

**Fig. 2.**

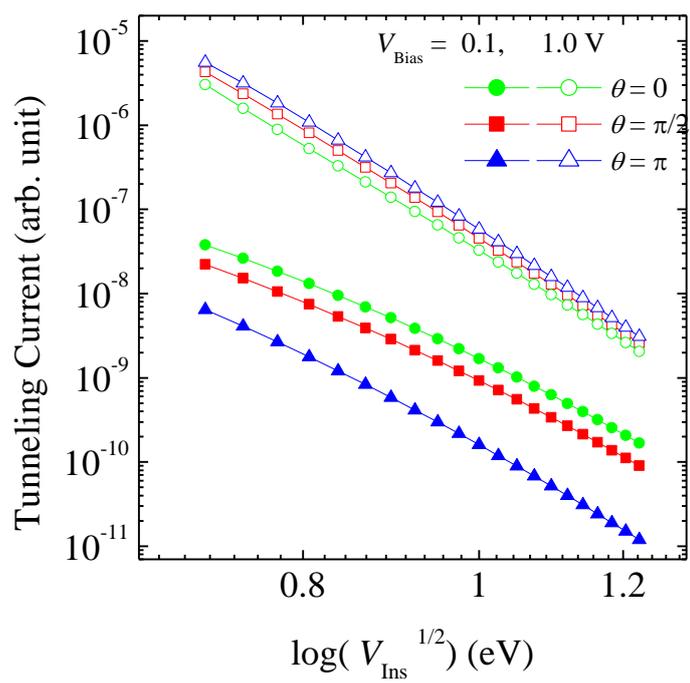

Fig. 2

**Fig. 3.**

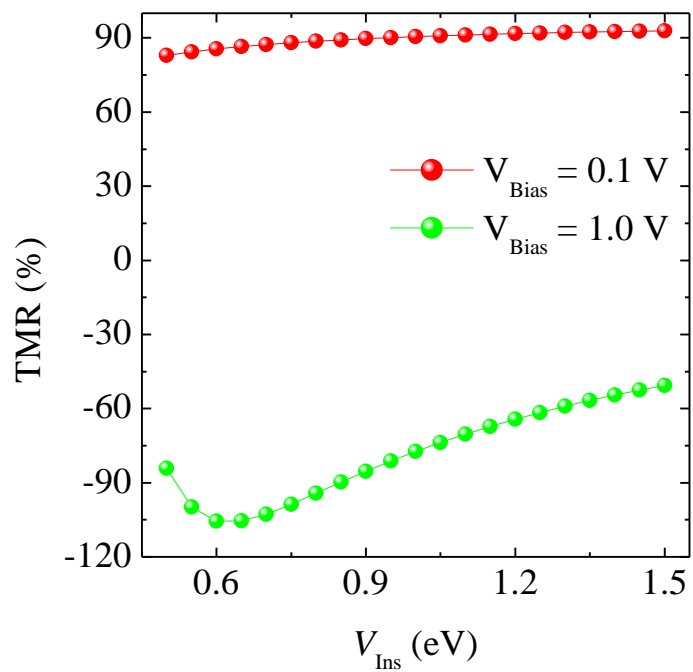

**Fig. 4.**

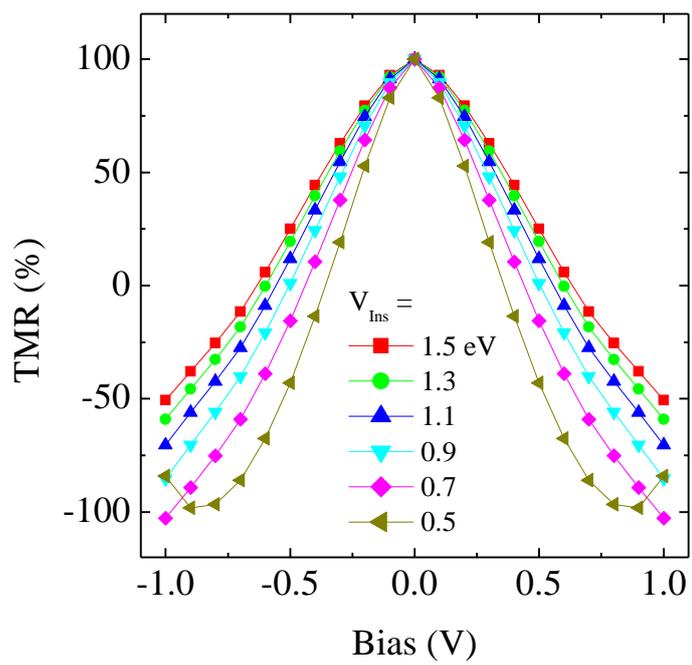

**Fig. 5 (a) and (b).**

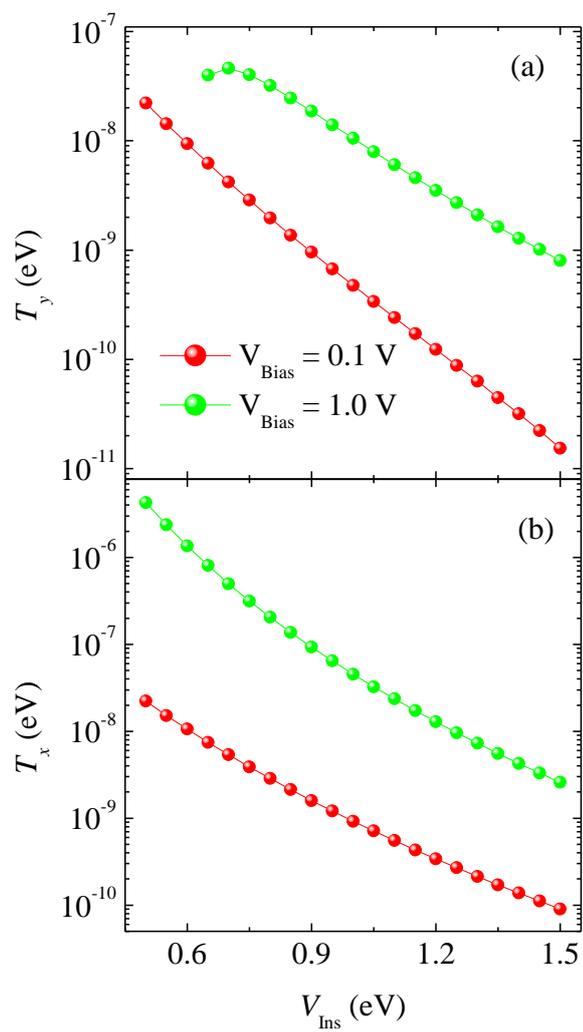

Fig. 5 (a), (b)

**Fig. 6 (a) and (b).**

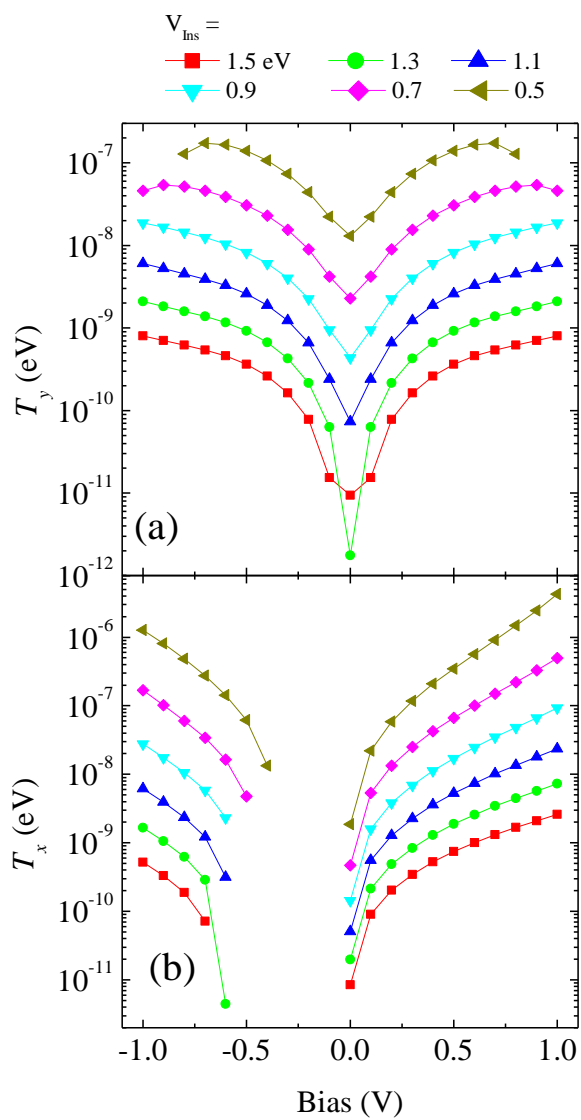

(a)

(b)

Fig. 6 (a), (b)

**Fig. 7.**

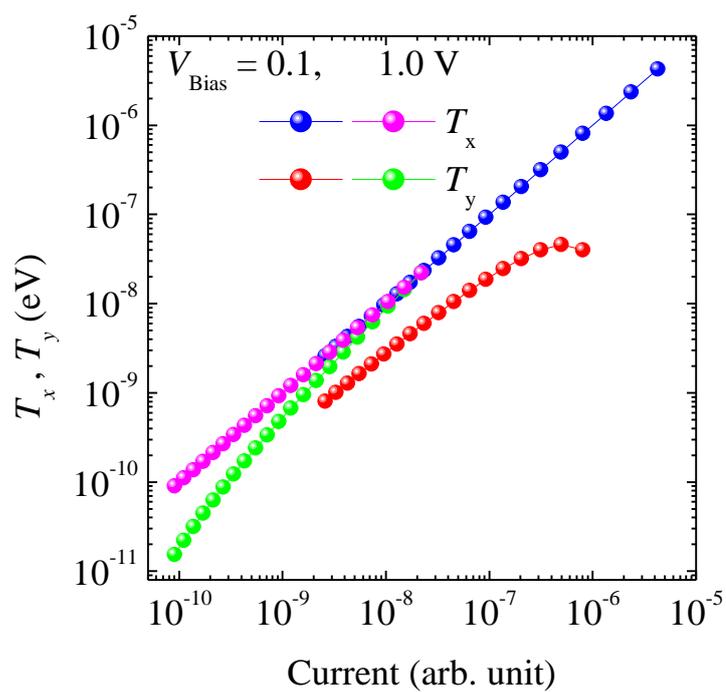

Fig. 7